\documentclass[prl,twocolumn,aps,floatfix,showpacs,tightenlines,superscriptaddress,amsmath,amssymb]{revtex4}
\usepackage{amsmath,amssymb,amsbsy,epsfig,color,graphicx,times,bm}
\usepackage{dcolumn}
\usepackage{bm}
\usepackage{psfrag}
\usepackage{mathbbol}

\newcommand{\ket}[1]{|#1\rangle}
\newcommand{\bra}[1]{\langle#1|}

\newcommand{\eq}[1]{Eq.~(\ref{#1})}

\newcommand{\fig}[1]{Fig.~\ref{#1}}

\begin{document}

\title{Long-distance entanglement and quantum teleportation in coupled cavity arrays}

\author{Salvatore M. Giampaolo}
\affiliation{Dipartimento di Matematica e Informatica,
Universit\`a degli Studi di Salerno, Via Ponte don Melillo,
I-84084 Fisciano (SA), Italy}
\affiliation{CNR-INFM Coherentia, and INFN Sezione di Napoli,
Gruppo collegato di Salerno, I-84084 Fisciano (SA), Italy}

\author{Fabrizio Illuminati}
\affiliation{Dipartimento di Matematica e Informatica,
Universit\`a degli Studi di Salerno, Via Ponte don Melillo,
I-84084 Fisciano (SA), Italy}
\affiliation{CNR-INFM Coherentia, and INFN Sezione di Napoli,
Gruppo collegato di Salerno, I-84084 Fisciano (SA), Italy}
\affiliation{ISI Foundation for
Scientific Interchange, Viale Settimio Severo 65, 00173 Torino,
Italy}
\affiliation{Corresponding author. Electronic address:
illuminati@sa.infn.it}

\pacs{03.65.Ca, 03.67.Mn, 73.43.Nq, 75.10.Jm}

\begin{abstract}
We introduce quantum spin models whose ground states allow for
sizeable entanglement between distant spins. We discuss how spin
models with global end-to-end entanglement realize quantum
teleportation channels with optimal compromise between scalability
and resilience to thermal decoherence, and can be implemented
straightforwardly in suitably engineered arrays of coupled optical cavities.
\end{abstract}

\date{October 21, 2009}

\maketitle

Experimental realizations of quantum communication \cite{Bennett}
and information \cite{Nielsen} protocols fall roughly in two
classes. The first one includes all-optical implementations, either
with single photons \cite{Bouwmeester,Boschi,Ursin}, or with
continuous variables \cite{Furusawa}. In the optical setting,
quantum communication is to a great extent decoherence-free and can
be easily carried out over long distances. However, the scalability
of all-optical devices is fundamentally limited, as the engineering
of strong interactions between photons poses formidable challenges.
The second class includes matter-based devices such as systems of
trapped ions \cite{Riebe,Barret}, superconducting quantum dots
\cite{Fattal}, and NMR-based devices \cite{Nielsen-2,Brassard}.
Matter-based implementations, that in principle are easily scalable,
suffer from hardly avoidable strong incoherent interactions with the
environment.
Moreover, typical schemes of quantum communication rely on properly
engineered direct interactions between microscopic constituents.
This requirement is due to the fact that in many-body systems and
spin chains, entanglement between individual constituents decays
very rapidly with the distance.

Crucial theoretical progress has been obtained with the recent
discovery that the ground state of particular classes of quantum
spin models with a finite correlation length can sustain
``long-distance entanglement'' (LDE), i.e. finite and large values
of the entanglement between distant spins even in the thermodynamic
limit \cite{Bologna,Salerno}. The LDE property, being a global,
non perturbative, ground-state feature is generated physically over
extremely short time scales (instantaneous, to all practical
purposes). Moreover, in a particular class of models that will be
introduced and discussed in the present work, LDE is enormously
stable against thermal fluctuations and decoherence. As we will
show, global LDE in these models is achieved by a minimal set of
local actions on the end-bond and near-end-bond couplings.
Therefore, these models can be realized in physical systems that
allow a sufficient degree of control, flexibility, and single-site
addressing. Quite recently, hybrid atom-optical systems of coupled
cavity arrays (CCAs) have been intensively studied in relation to
their ability to realize/simulate collective phenomena typical of
strongly correlated systems
\cite{Hartmann,Greentree,Angelakis,Rossini,Hartmann-2}. In fact, the
extremely high controllability, the straightforward addressability
of single constituents, and the great degree of flexibility in their
geometric design \cite{Hartmann-3}, make CCAs strong candidates for
the realization of extended communication networks and scalable
computational devices.

In this work we investigate quantum spin systems and schemes for the
realization of long-distance quantum teleportation based on the phenomenon
of LDE, and we illustrate their experimental feasibility in suitably
engineered CCAs. We first introduce a class of spin models with open
ends and suitably defined end-bond and near-end-bond interactions,
and we discuss how the ground-state structure of these models
sustains global LDE and allows for long-distance and high-fidelity
end-to-end teleportation, even at moderately high temperatures. We
then show how these quantum spin channels, that conjugate
scalability and resilience to decoherence, can be implemented by
open-end, one-dimensional CCAs with properly engineered local
couplings at each end. Finally, by exploiting the high degree of
control and flexibility of CCAs, we introduce quasi-deterministic
schemes of teleportation with high success rates without direct
projection on Bell states and Bell measurements, thus overcoming one
of the major difficulties that typically beset matter-based devices.

Let us first consider quantum spin models defined on a
one-dimensional lattice of length $N$, with open ends, and general
site-dependent nearest-neighbor interactions of the $XX$ type:
\begin{equation}
\label{Hamiltoniana di spin}
H_s \, = \, - \sum_{k=1}^{N-1}J_{k} (S^x_k S^x_{k+1}+S^y_k S^y_{k+1}) \; ,
\end{equation}
where ${J_k}$ is the set of the $N-1$ nearest-neighbor couplings and
$S^\alpha_k$ denotes the Spin-1/2 operators at site $k$
($\alpha=x,y$). The pure $XX$ limit is recovered when $J_k = J$,
$\forall k$. The model is exactly solvable by Jordan-Wigner
diagonalization, both for chains of finite size and in the
thermodynamic limit \cite{Lieb}. For arbitrary site-dependent
couplings, the associated models are still exactly solvable by a
straightforward extension of the techniques developed in Ref.
\cite{Lieb}, albeit, in general, only numerically
\cite{Salerno,unpublished,wojcik05}. For these models, the
phenomenon of perfect ground-state LDE (maximal end-to-end
concurrence independent of the size of the chain) sets on when the
Hamiltonian $H_s$ is dimerized, with perfectly alternated weak and
strong couplings: $J_k \equiv J_{odd}$ ($k=1,3,...,N-1$), $J_k
\equiv J_{even}$ ($k=2,4,...,N-2$), and $J_{odd} \ll J_{even}$
\cite{Salerno}. Normalizing all couplings by $J_{even}$, and
renaming the rescaled odd coupling: $J_{odd}/J_{even} \equiv
\lambda$, the condition for perfect LDE reads $\lambda \ll 1$
\cite{Salerno}. Ground-state quasi-perfect LDE (maximal end-to-end
concurrence slowly decaying with the size of the chain) is realized
by models with uniform bulk (b) interactions and weak end bonds
(eb): $J_2=J_3=...=J_{N-2} \equiv J_{b}$, $J_{1}=J_{N-1} \equiv
J_{eb}$. Rescaling all couplings by $J_{b}$ (with $J_{eb}/J_{b}
\equiv \lambda$), the condition for quasi-perfect LDE reads $\lambda
\ll 1$ \cite{Salerno}. In the first instance ({\it dimerized $XX$
model}) the energy gap closes exponentially with the size of the
system, making this system useless for realistic applications at
finite temperature. In the second instance (the {\it $\lambda$
model}) the gap closes with an algebraic power law, but useful
amounts of LDE can survive only at extremely low temperatures,
unreachable at present and in the immediately foreseeable future
\cite{Salerno}. Here we introduce a {\it $\lambda$-$\mu$ model} that
realizes an optimal compromise between the requirements of strong
LDE in a system of large size, robustness of LDE at moderately high
temperatures, and ease of realistic implementations. The
$\lambda$-$\mu$ model is defined by Hamiltonian (\ref{Hamiltoniana
di spin}) with uniform bulk interactions and alternating weak end
bonds (eb) and strong near-end bond (neb) interactions:
$J_3=J_4=...=J_{N-3} \equiv J_{b}$, $J_{2}=J_{N-2} \equiv J_{neb}$,
and $J_1=J_{N-1} \equiv J_{eb}$. Normalizing all interactions by
$J_{b}$, and redefining the scaled couplings $J_{eb}/J_{b} \equiv
\lambda$ and $J_{neb}/J_{b} \equiv \mu$, the condition for an
optimized end-to-end LDE is $\lambda \ll 1 \ll \mu$.
\begin{figure}[h]
\centering
\begin{tabular}{c}
\epsfig{file=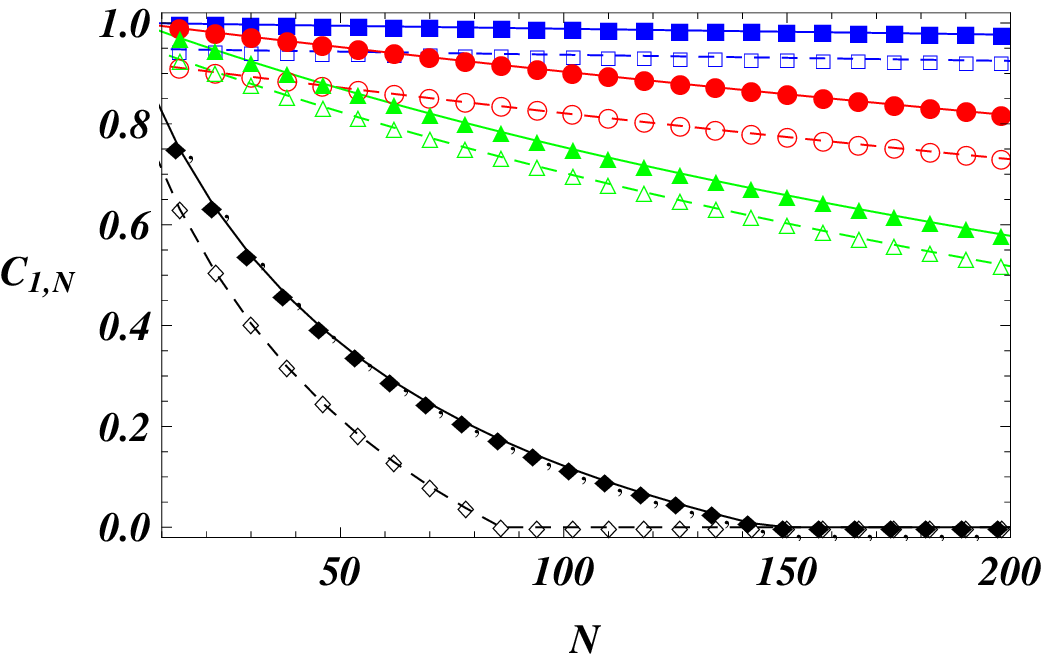,width=0.75\linewidth,clip=} \\
\epsfig{file=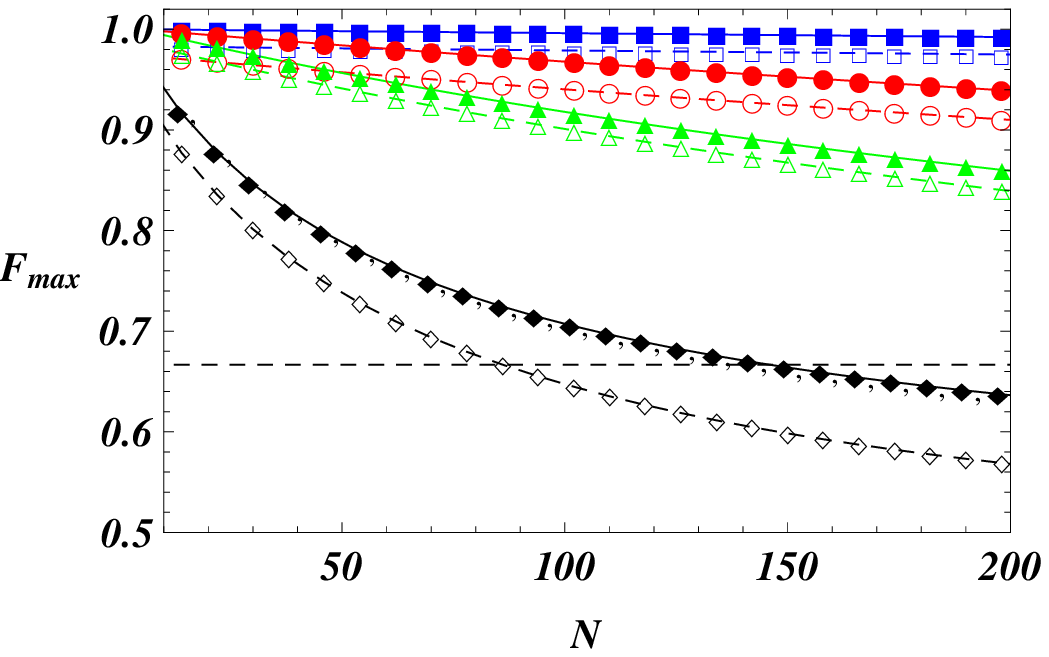,width=0.75\linewidth,clip=}
\end{tabular}
\caption{(Color online) Upper panel, lines with full symbols:
Concurrence ${\cal{C}}_{1N}$ between the end points of a $\lambda$-$\mu$ chain as a function of
the length $N$, for different values of $\lambda$ and $\mu$, at zero temperature.
From top to bottom: Line with full boxes: $\lambda = 0.15$, $\mu = 7.0$.
Full circles: $\lambda = 0.2$, $\mu = 5.0$. Full triangles:
$\lambda = 0.4$, $\mu = 3.0$. Full diamonds: ${\cal{C}}_{1N}$ of a $\lambda$
spin chain (i.e. with $\mu = 1$) for $\lambda = 0.2$. Upper panel, lines with empty symbols:
${\cal{C}}_{1N}$ as a function of $N$ at different reduced temperatures $T/J_{b}$ for
the same sets of values of $\lambda$ and $\mu$ as for the corresponding lines with
full symbols. From top to bottom: Line with empty boxes: $T/J_{b} = 0.0005$.
Empty circles: $T/J_{b} = 0.0001$. Empty triangles: $T/J_{b} = 0.001$.
Empty diamonds: ${\cal{C}}_{1N}$ of the corresponding $\lambda$ spin chain at
$T/J_{b} = 0.0006$. Lower Panel: The maximal fidelity of teleportation ${\cal{F}}_{max}$
\cite{Horodecki} between the end points of the $\lambda$-$\mu$ chain as a function of $N$,
at zero and finite temperature, for the same set of values reported in the
upper panel. In the case of the $\lambda$ model (lines with diamonds),
${\cal{F}}_{max}$ sinks below the classical threshold $2/3$
(horizontal dashed line), for the corresponding vanishing values of
the end-to-end concurrence (See upper panel). All quantities being plotted
are dimensionless.} \label{Concurrence
and Fidelity}
\end{figure}

The LDE properties of the $\lambda$-$\mu$ chain, at zero and finite
temperature, are reported in the upper panel of Fig.
\ref{Concurrence and Fidelity}, where the end-to-end concurrence is
plotted as a function of the size of the chain. In the lower panel
of Fig. \ref{Concurrence and Fidelity} we report the behavior of the
corresponding maximal fidelity of teleportation ${\cal{F}}_{max}$,
which, in the case of nonvanishing spin-spin concurrence, is a
monotonic function of the pairwise end-to-end entanglement
\cite{Horodecki,Salerno}. Fig. \ref{Concurrence and Fidelity} shows
that interacting spin systems of the $\lambda$-$\mu$ type conjugate
efficiently resilience to decoherence and scalability, as further
confirmed by comparing with the performance of systems of the $\lambda$ type.

\begin{figure}[t]
\includegraphics[width=6.cm]{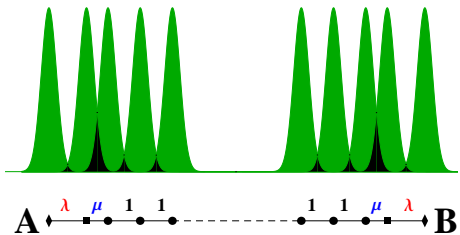}
\caption{(Color online) Scheme of a CCA realizing the
$\lambda$-$\mu$ $XX$ model. Dark green (grey in black and white
print): Overlap of the wave functions
associated to each site of the array. The two next-to-end sites
(boxes) are symmetrically displaced towards their
neighbors in the bulk (circles). The overlap (black area)
between the wave functions of these two cavities and their neighbors
in the bulk is thus larger than the would-be reference (unit) one in
an equispaced CCA. Viceversa, the overlap between the end sites of
the array (diamonds) and the next-to-end sites is reduced proportionally
compared to an equispaced array.} \label{CCA}
\end{figure}

Let us now consider a linear CCA with open ends, consisting of $N$
cavities. The dynamics of a single constituent of the array doped
with a single two-level atom is well described by the
Jaynes-Cummings Hamiltonian
\begin{equation}
\label{Hamiltoniana locale}
  H_k = \omega a^\dagger_k a_k+\omega' \ket{e_k}\bra{e_k} +
  g  a^\dagger_k \ket{g_k}\bra{e_k} + g\ket{e_k}\bra{a_k} a_k \, ,
\end{equation}
where $a_k$ ($a^\dagger_k$) is the annihilation (creation) operator
of photons with energy $\omega$ in the $k$-th cavity, $\ket{g_k}$
and $\ket{e_k}$ are respectively the ground and excited atomic
states, separated by the gap $\omega'$, and $g$ is the photon-atom
coupling strength. The local Hamiltonian \eq{Hamiltoniana locale} is
immediately diagonalized in the basis of dressed photonic and atomic
excitations (polaritons):
\begin{eqnarray}\label{autostati}
 \ket{\emptyset_k} & = & \ket{g_k}\ket {0_k} \, ;\\
 \ket{n+_k} & = & \cos\theta_n \ket{g_k} \ket{n_k} + \sin \theta_n
 \ket{e_k}\ket{(n-1)_k} \; \; n \ge 1\,  ; \nonumber \\
 \ket{n-_k} & = & \sin\theta_n \ket{g_k}\ket{n_k} - \cos \theta_n
 \ket{e_k}\ket{(n-1)_k} \; \; n \ge 1\, , \nonumber
\end{eqnarray}
where $\theta_n$ is given by $\tan 2 \theta_n=-g\sqrt{n}/\Delta$ and
$\Delta=\omega'-\omega$ is the atom-light detuning. Each polariton
is characterized by an energy equal to
\begin{figure}[t]
\includegraphics[width=6. cm]{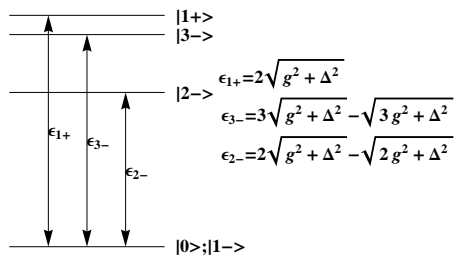}
\caption{Representation of the energy levels for a cavity with
$\omega = \sqrt{ g^2+\Delta^2}$. The ground state is two-fold
degenerate, and the energy gap $\varepsilon_{2-}$ prevents the
occupancy of the higher energy levels, thus realizing an effective
two-level system.} \label{livelli}
\end{figure}
\begin{equation}\label{autoenergie}
 \varepsilon_0 = 0 ; \; \; \; \; \; \;
 \varepsilon_{n\pm} = n \omega \pm \sqrt{n g^2+\Delta^2} \,.
\end{equation}
When $\omega = \sqrt{ g^2+\Delta^2}$ the ground state of
\eq{Hamiltoniana locale} becomes two-fold degenerate, resulting in a
superposition of $\ket{\emptyset_k}$ and $\ket{1-_k}$, see
\fig{livelli}. If both the atom-cavity interaction energy and the
working temperature are small compared to $\varepsilon_{2-}=2 \sqrt{
g^2+\Delta^2} - \sqrt{2 g^2+\Delta^2}$, one may neglect all the
local polaritonic states but $\ket{\emptyset_k}$ and $\ket{1-_k}$.
This situation defines a local two-level system. Adjacent cavities
can be easily coupled either by photon hopping or via wave guides of
different dielectric and conducting properties. The wave function
overlap between adjacent cavities introduces the associated
tunneling elements, so that the total Hamiltonian of the CCA is
\begin{equation}
\label{Hamiltoniana totale}
H_{cca} = \sum_k^N H_k  - \sum_k^{N-1} J_{k} (a^\dagger_k a_{k+1} + a^\dagger_{k+1} a_{k}) \; .
\end{equation}
Each hopping amplitude $J_k$ depends strongly on both the geometry
of the cavities and the actual overlap between adjacent cavities. If
the maximum value among all the couplings $\{ J_k \}$ is much below
the energy of the first excited state: $\max\{J_k\}\ll
\varepsilon_{2-}$, then the total Hamiltonian \eq{Hamiltoniana
totale} can be mapped in a spin-1/2 model of the $XX$ type with
site-dependent couplings of the form \eq{Hamiltoniana di spin},
where the state $\ket{\emptyset_k}$ ($\ket{1-_k}$) plays the role of
$\ket{\downarrow_k}$  ($\ket{\uparrow_k}$). The mapping to an
open-end $\lambda$-$\mu$ linear spin chain is then realized, e.g.,
by simply tuning the distance between the end- and next-to-end sites
of the CCA, as showed in \fig{CCA}.
The $\lambda$-$\mu$ chain is thus realized by placing the second
and the $(N-1)-th$ cavities slightly off their would-be positions
in an equispaced CCA. This shift lowers the
coupling between the two cavities and those at the end-points of the chain:
$J_1/J_b=J_{N-1}/J_b=\lambda < 1$, where $J_b$ is the
uniform nearest-neighbor coupling in the bulk. Viceversa, it
increases the coupling between the two next-to-end sites and
their nearest neighbors in the bulk: $J_2/J_b=J_{N-2}/J_b=\mu > 1$.

\begin{figure}[t]
{\includegraphics[width=3.1cm]{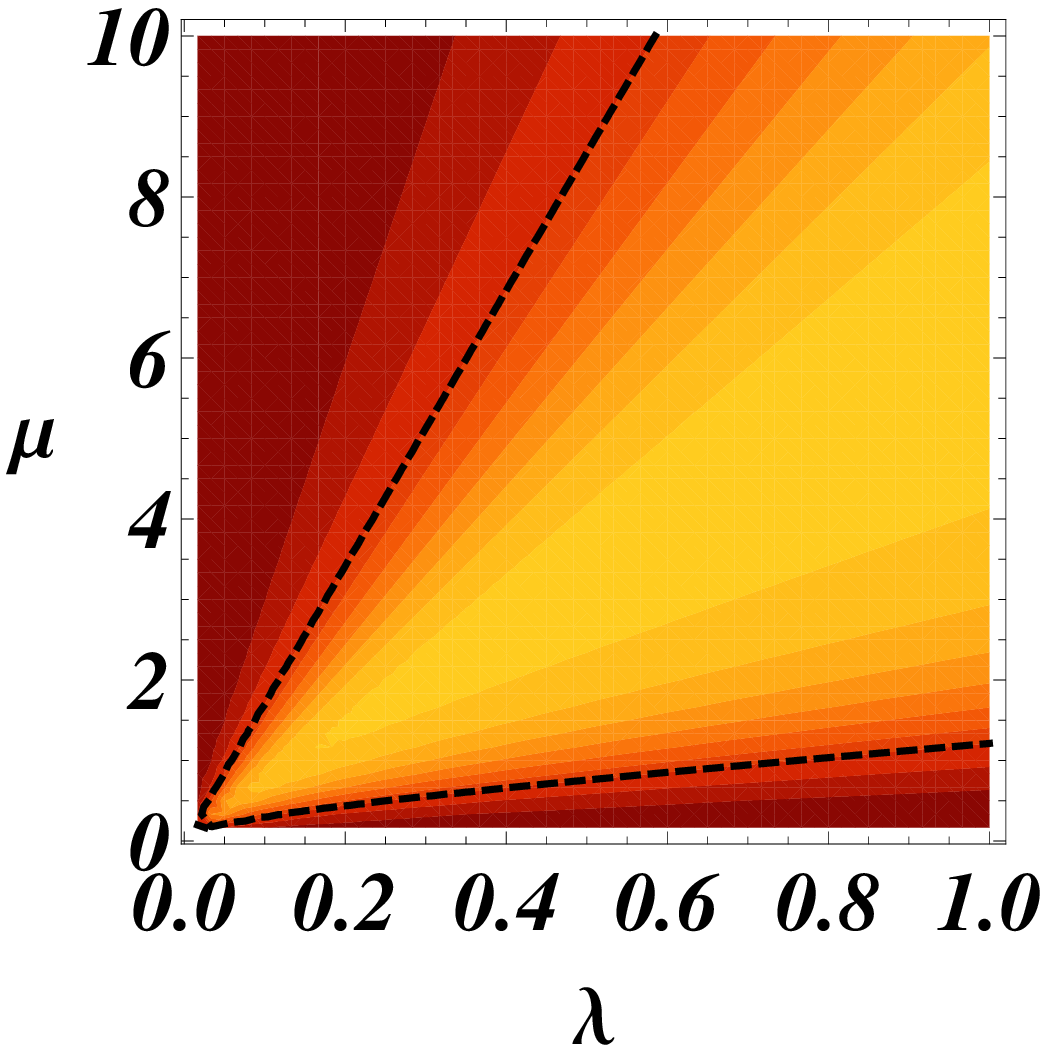}
\includegraphics[width=3.1cm]{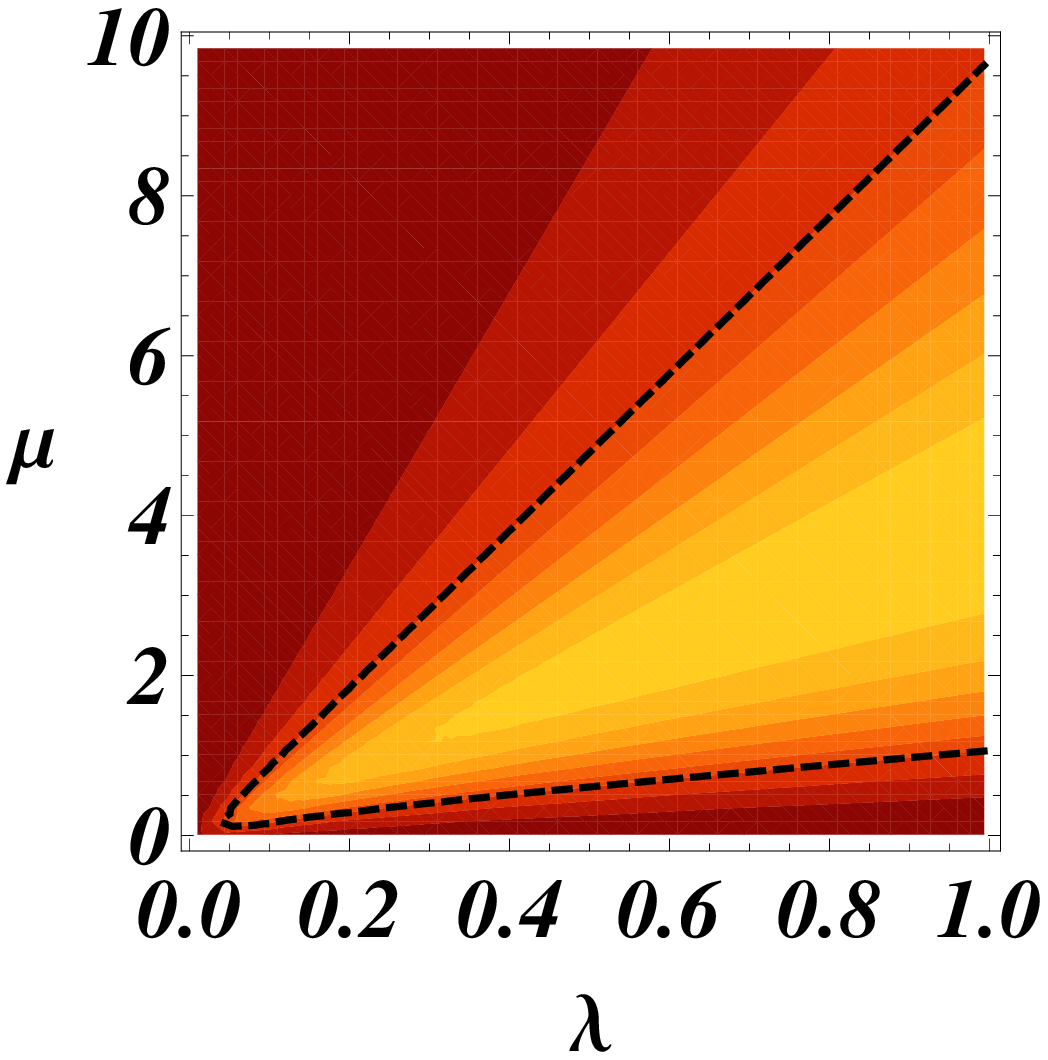}\hspace{.2cm}
\includegraphics[width=0.7cm]{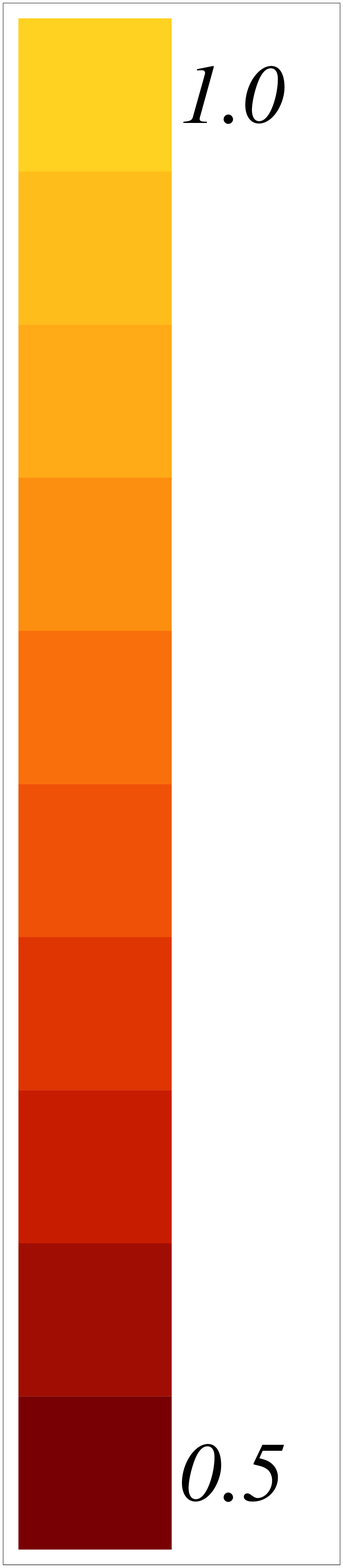}}
\caption{(Color online) Fidelity of teleportation $F_{max}$ in a
$\lambda$-$\mu$ configuration, by exact numerical diagonalization,
for a CCA of $N=12$ cavities, as a function of the couplings
$\lambda=J_1/J_b$ and $\mu=J_2/J_b$ and at different temperatures
$T/J_b$. Left panel: $T/J_b = 0.005$. Right panel: $T/J_b = 0.01$.
$F_{max}$ varies between $0.5$ (dark red in color, dark grey in black and
white print) and $1.0$ (light orange in color, light grey in black and
white print). Dashed lines: Classical threshold $F_{max}^c = 2/3$. All
quantities being plotted are dimensionless.}
\label{Maxfidelity-1}
\end{figure}

We now proceed to illustrate that CCAs in the $\lambda$-$\mu$
configuration allow for long-distance and high-fidelity quantum
communication in fully realistic conditions at moderately high
temperatures. In Fig. \ref{Maxfidelity-1} we report the fidelity of
teleportation $F_{max}$ as a function of the reduced couplings
$\lambda$ and $\mu$ for different temperatures.
Remarkably, \fig{Maxfidelity-1} shows the existence of a rather high
{\it critical} temperature of teleportation for CCAs in the
$\lambda$-$\mu$ configuration. The region of the physical parameters
compatible with a nonclassical fidelity $F_{max} > 2/3$ is
progressively reduced with increasing temperature, until it
disappears at $T_c \approx 0.13 J_b$. Similar behaviors are observed
for longer CCAs, with $T_{c}$ slowly decreasing with the length of
the array. For instance, for an array of $N=36$ cavities in the
$\lambda$-$\mu$ configuration, the critical temperature of
transition to {\it bona fide} quantum teleportation is $T_c \approx
0.11 J_b$.

A formidable obstacle to the concrete realization of working quantum
teleportation devices is performing the projection over a Bell
state, in order for the sender to teleport a quantum state
faithfully to the receiver. In fact, in the framework of condensed
matter there hardly exist quantities, easily available in current
and foreseeable experiments, that admit as eigenstates any two-qubit
Bell states. In the following, we will illustrate a simple and
concrete scheme for long-distance, high-fidelity quantum
teleportation in $\lambda$-$\mu$ CCAs that realizes Bell-state
projections indirectly, by matching together free evolutions and
local measurements of easily controllable experimental quantities
\cite{Zheng,Ye,Cardoso}. We first illustrate it in the simplest case
of two cavities at zero temperature, with the first cavity
accessible by the sender and the second one by the receiver. The
sender has access also to a third cavity, the "0" cavity, that is
decoupled from the rest of the chain and stores
the state to be teleported $\ket{\varphi}=\alpha
\ket{\uparrow_0}+\beta \ket{\downarrow_0}$. The decoupling
of the $0$-th cavity is achieved, e. g., by
removing the degeneracy among $\ket{0}_0$ and $\ket{1-}_0$ and
taking $|\varepsilon_0-\varepsilon_{1-}|\gg J_0$.
The total system is initially in the state
\begin{equation}\label{groundstate}
\ket{\Psi(0)}=\frac{1}{\sqrt{2}}(\alpha \ket{\uparrow_0}+\beta
\ket{\downarrow_0}) (\ket{\uparrow_1} \ket{\downarrow_2}
+\ket{\downarrow_1}\ket{\uparrow_2}).
\end{equation}
At $t=0$ the state begins to evolve and, if $J_0 \gg J_1$, one
has:
\begin{eqnarray}\label{initialstate}
\ket{\Psi(t)}&=&\frac{1}{\sqrt{2}}\left[ \alpha
\ket{\uparrow_0}\ket{\uparrow_1}\ket{\downarrow_2}+\beta
\ket{\downarrow_0}\ket{\downarrow_1}\ket{\uparrow_2} \right.
\\
& & \ket{\uparrow_0}\ket{\downarrow_1} (\alpha \cos(J_0 t)
\ket{\uparrow_2}- i \beta \sin(J_0 t) \ket{\downarrow_2} )\nonumber \\
& &\left. \ket{\downarrow_0}\ket{\uparrow_1} (-i \alpha \sin(J_0 t)
\ket{\uparrow_2}+ \beta \cos(J_0 t) \ket{\downarrow_2}) \right] .
\nonumber
\end{eqnarray}
If at time $t=\pi/(4 J_0)$ Alice measures the local magnetizations
($S_0^z$, $S_1^z$) in the first two cavities, she will find with
probability $1/2$ that the teleported state is the image of
$\ket{\varphi}$ under a local rotation. The value $1/2$ for the
probability stems from the fact that any simultaneous eigenstate of
$S_0^z$ and $S_1^z$ can be obtained with equal probability but one
may discard the case in which the total magnetization is equal to
$\pm1$. Realizing a local rotation of $\pm \pi/2$ around $S^z_2$,
with the sign depending on the result of the measurement that the
sender communicates classically to the receiver, the latter recovers
the original state $\ket{\varphi}$ with unit fidelity.



\begin{figure}[t]
\includegraphics[width=6. cm]{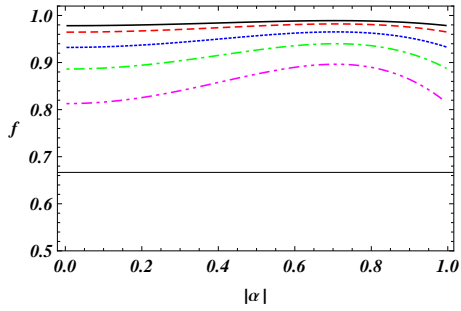}
\caption{(Color online) Fidelity of teleportation $f$ in a $\lambda$-$\mu$
CCA channel of $N=12$ cavities, for a generic input
$\ket{\varphi}=\alpha \ket{\uparrow_0}+\beta \ket{\downarrow_0}$, as
a function of $|\alpha|$ at different temperatures. From top to
bottom: Solid line: $T = 0.001 J_b $. Dashed line: $T = 0.003 J_b$.
Dotted line: $T = 0.004 J_b$. Dot-dashed line: $T = 0.005 J_b$. Double-dot-dashed
line: $T = 0.007 J_b$. Here $\lambda = 0.5$, $\mu = 4.0$, and $\nu = 50$. Horizontal
solid line: Classical threshold $f_{c}=2/3$. All quantities being plotted are
dimensionless.}
\label{fidelity}

\end{figure}

The simple protocol described above can be immediately extended to
$\lambda$-$\mu$ CCAs of any size, at finite temperature, and
removing the constraint $J_1 \ll J_0$. By resorting again to exact
diagonalization, in \fig{fidelity} we report the behavior of the
fidelity of teleportation $f$, as a function of $|\alpha|$ of the
state $\ket{\varphi}$, in the case of an array of $N=12$ cavities,
with $\nu \equiv J_0/J_b = 50$ and for different temperatures. Also
in the non-ideal case the teleportation protocol has probability 1/2
of success. The fidelity depends on the input state, with a maximum
for inputs with $|\alpha|=|\beta|=1/\sqrt{2}$ and a minimum for
inputs with $|\alpha|=0.1$. The fidelity remains above 0.95 for all
values of $|\alpha|$ at moderately low temperatures ($T \simeq
10^{-3}J_{b}$).



In conclusion, we have introduced a class of quantum spin models
characterized by a non-perturbative, ground-state (quasi) long-distance
entanglement strongly resilient to thermal decoherence, that can be
efficiently realized with a minimal set of local actions on the end sites
of open CCAs. We have showed that these systems allow for a simple
probabilistic protocol of long-distance, high-fidelity quantum
teleportation that yields a high rate of success without direct Bell
measurements and projections over Bell states.
We acknowledge financial support from the EC under the FP7 STREP
Project HIP, Grant Agreement n. 221889, from MIUR under the FARB
fund, and from INFN under Iniziativa Specifica PG 62. F. I. acknowledges
support from the ISI Foundation for Scientific Interchange.


\begin{thebibliography}{99}


\bibitem{Bennett} C. H. Bennett, G. Brassard, C. Cr\'epeau, R. Jozsa, A. Peres,
and W. K. Wootters, Phys. Rev. Lett. {\bf 70}, 1895 (1993).

\bibitem{Nielsen} M. A. Nielsen and I. L. Chuang, {\em Quantum Computation and
Quantum Information} (Cambridge University Press, Cambridge, 2000).

\bibitem{Bouwmeester} D. Bouwmeester, J.-W. Pan, K. Mattle, M. Eibl, H.
Weinfurter, and A. Zeilinger, Nature {\bf 390}, 575 (1997).

\bibitem{Boschi} D. Boschi, S. Branca, F. De Martini, L. Hardy, and S. Popescu,
Phys. Rev. Lett. {\bf 80}, 1121 (1998);

\bibitem{Ursin} R. Ursin {\it et al.}, Nature {\bf 430}, 849 (2004).

\bibitem{Furusawa} H. Yonezawa, T. Aoki, and A. Furusawa, Nature {\bf 431}, 430 (2004).


\bibitem{Riebe} M. Riebe {\it et al.}, Nature {\bf 429}, 734 (2004).

\bibitem{Barret} M. D. Barrett {\it et al.}, Nature {\bf 429}, 737 (2004).

\bibitem{Fattal} D. Fattal, E. Diamanti, K. Inoue, and Y. Yamamoto,
Phys. Rev. Lett. {\bf 92}, 037904 (2004).

\bibitem{Nielsen-2} M. A. Nielsen, E. Knill, and R. Laflamme, Nature {\bf 395}, (1998).

\bibitem{Brassard} G. Brassard, S. L. Braunstein, and R. Cleve, Physica D {\bf 120}, 43 (1998).





\bibitem{Bologna} L. Campos Venuti, C. Degli Esposti Boschi, and
M. Roncaglia, Phys. Rev. Lett. {\bf 96}, 247206 (2006);
Phys. Rev. Lett. {\bf 99}, 060401 (2007).

\bibitem{Salerno} L. Campos Venuti, S. M. Giampaolo, F. Illuminati, and P. Zanardi,
Phys. Rev. A {\bf 76}, 052328 (2007).

\bibitem{Hartmann} M. J. Hartmann, F. G. S. L. Brand$\tilde{\mathrm{a}}$o,
and M. B. Plenio, Nature Phys. {\bf 2}, 849 (2006).

\bibitem{Greentree} A. D. Greentree, C. Tahan, J. H. Cole, and L. C. L. Hollenberg, Nature Phys.
{\bf 2}, 856 (2006).

\bibitem{Angelakis} D. G. Angelakis, M. F. Santos, and S. Bose, Phys. Rev. A {\bf 76},
031805(R) (2007).

\bibitem{Rossini} D. Rossini and R. Fazio, Phys. Rev. Lett. {\bf 99}, 186401 (2007).

\bibitem{Hartmann-2} M. J. Hartmann, F. G. S. L. Brand$\tilde{\mathrm{a}}$o,
and M. B. Plenio, Phys. Rev. Lett. {\bf 99}, 160501 (2007).

\bibitem{Hartmann-3} M. J. Hartmann, F. G. S. L. Brand$\tilde{\mathrm{a}}$o,
and M. B. Plenio, Laser \& Photon. Rev. {\bf 2}, 527 (2008).

\bibitem{Lieb} E. Lieb, T. Schultz, and D. Mattis, Ann. Phys. (N.Y.)
{\bf 16}, 407 (1961).

\bibitem{unpublished} S. M. Giampaolo and F. Illuminati, unpublished.

\bibitem{wojcik05} A. Wojcik {\it et al.}, Phys. Rev. A {\bf 72}, 034303 (2005).

\bibitem{Horodecki} M. Horodecki, P. Horodecki, and R. Horodecki,
Phys. Rev. A. {\bf 60}, 1888 (1999); P. Badziag, M. Horodecki, P.
Horodecki, and R. Horodecki, Phys. Rev. A  {\bf 62}, 012311 (2000)


\bibitem{Zheng} S.-B. Zheng, Phys. Rev. A {\bf 69}, 064302 (2004).

\bibitem{Ye} L. Ye and G-C. Guo, Phys. Rev. A {\bf 70}, 054303 (2004).

\bibitem{Cardoso} W. B. Cardoso, A. T. Avelar, B. Baseia, and N. G. de Almeida, Phys. Rev. A {\bf 72},
045802 (2005).







\end{thebibliography}
\end{document}